\documentclass[twocolumn,pra,showpacs,floatfix]{revtex4}
\usepackage{bm}                      
\usepackage{graphicx}                
\usepackage{dcolumn}                 
\usepackage{amsmath}
\usepackage{amsfonts}

\newcommand*{\cent}[1]{\multicolumn{1}{c}{$#1$}}
\newcolumntype{w}[1]{D{.}{.}{#1}}
\newcommand{\ds}{\displaystyle}
\newcommand{\br}{\vec{r}}
\newcommand{\bs}{\vec{s}}
\newcommand{\phiel}{\phi_\mathrm{el}}
\newcommand{\Hinv}{\frac{1}{{\cal E}_{\rm el}-H_{\rm el}}}
\newcommand{\Hinvp}{\frac{1}{({\cal E}_{\rm el}-H_{\rm el})'}}
\newcommand{\nrs}{\nabla_{\!R}^2}
\newcommand{\nr}[1]{{\vec \nabla}_{\!R}^{#1}}

\begin{document}

\title{Nonadiabatic corrections to rovibrational levels of H$_2$}

\author{Krzysztof Pachucki}
\email{krp@fuw.edu.pl}
\affiliation{Institute of Theoretical Physics,
             University of Warsaw, Ho{\.z}a 69, 00-681~Warsaw, Poland }
\author{Jacek Komasa}
\email{komasa@man.poznan.pl}
\affiliation{Faculty of Chemistry, 
        A.~Mickiewicz University, Grunwaldzka 6, 60-780~Pozna\'n, Poland }

\date{\today}

\begin{abstract}
The leading nonadiabatic corrections to rovibrational levels of a diatomic
molecule are expressed in terms of three functions of internuclear distance: 
corrections to the adiabatic potential, the effective nuclear mass, 
and the effective moment of inertia. The resulting radial Schr\"odinger equation 
for nuclear motion is solved numerically yielding accurate nonadiabatic 
energies for all rovibrational levels of H$_2$ molecule in excellent 
agreement with previous calculations by Wolniewicz.
\end{abstract}

\maketitle

\section{Introduction}

In the fully nonadiabatic approach the total nonrelativistic energy 
of a molecular state is obtained by solving the Schr{\"o}dinger equation with  
kinetic energy of electrons and of nuclei on the same footing.
This approach has been applied to vibrational states
of several small diatomic molecules \cite{KW63,BA03,CA05,BA05,SKBMA08a}.
Much more commonly though, the total energy of a molecular state is obtained
in a two-step procedure based on the Born-Oppenheimer (BO) approximation 
\cite{BO27,BH55} in which a separation of electronic and nuclear motion is 
assumed. Namely, in the first step, the electronic Schr{\"o}dinger equation with 
the clamped nuclei Hamiltonian is solved for different nuclear configurations 
yielding the electronic energy as a function of the nuclear coordinates. 
This function,  called the potential energy surface (PES),
serves as a potential for the motion of nuclei in the nuclear 
Schr{\"o}dinger equation. Solving this equation in the second step 
yields the total rovibronic energy of the system. 

The accuracy of theoretical predictions for molecular states,
limited due to the BO approximation, can be increased by including
the adiabatic, relativistic, and radiative corrections without
renouncing the notion of PES. Further increase in the accuracy requires
the nonadiabatic effects to be taken into account. A desirable way of inclusion
of these effects is in terms of a geometry dependent  function, 
which can be added to PES in the same manner as all the other
corrections. On one hand, the nuclear Schr{\"o}dinger equation, when solved with 
such a potential, gives the molecular energy levels with spectroscopic 
precision. On the other, the notion of the PES is 
preserved with all its advantages. 

Several, more or less successful attempts to construct such a nonadiabatic 
correction function for a diatomic molecule can be found in literature
\cite{HA66,BM77,HO98,Kut07,JK08}. Bunker and Moss  have derived \cite{BM77}, 
in the second order of the perturbative expansion, an effective  nuclear 
Hamiltonian for the ground electronic state of diatomic molecules, 
in terms of the nonadiabatic potential, vibrational and rotational masses.
Assuming constant vibrational and rotational masses
and neglecting the nonadiabatic potential, they obtained values 
of these effective masses for H$_2$ and D$_2$ by fitting to the experimental data.
Later Schwenke in \cite{SCH01} used the Bunker and Moss \cite{BM77,BM80} 
effective Hamiltonian to perform ab initio calculations
of nonadiabatic corrections for H$_2$ and H$_2$O.
His results for purely vibrational spectrum of H$_2$ differ from that of Wolniewicz
\cite{Wol95} by about 20\% due to the inaccurate numerical representation 
of the wave function and, what we demonstrate in this work, due to the approximate 
second order nonadiabatic potential of Bunker and Moss \cite{BM77}. 
The accuracy of Schwenke calculations \cite{SCH01} for H$_2$O is probably
not higher, but clearly demonstrates wide applicability of the perturbative approach.
Very recently Kutzelnigg {\em et al.} \cite{Kut07,JK08} performed
simplified calculations of the nonadiabatic correction and both effective masses
as functions of the internuclear distance in H$_2^+$ and H$_2$. 
In our recent work \cite{PK08b} we have introduced nonadiabatic perturbative 
theory and derived formulae for the leading nonadiabatic corrections
to energies and wave functions. The formula for the nonadiabatic energy from
that work, although apparently different, is in fact equivalent to that of Bunker 
and Moss \cite{BM77}. Our results for rotationless vibrational states 
have been obtained as the expectation value of nonadiabatic corrections  
with the adiabatic wave function. Although numerically accurate,
due to the neglected third order nonadiabatic corrections [see Eq.~(\ref{Ed3E})],
our results differed by about 2\%  from the previous calculations 
by Wolniewicz \cite{Wol95} and by Stanke {\em et al.} \cite{SKBMA08b}. 

In this paper, the nonadiabatic perturbation theory
has been extended in two directions. Firstly, we generalize
the previous derivation to rotational states.
Secondly, we include the previously missing third order correction, 
which has proved significant.
Moreover, we present a rigorous formulation of the nonadiabatic 
perturbative theory and include the numerical example of the H$_2$ molecule. 
This can be extended to any diatomic molecule 
and potentially to an arbitrarily large molecule.
We derive formulae valid to all orders,
present the leading corrections of order $\mathcal{O}(\mu_n^{-2})$,
and express them in terms of the nonadiabatic correction to the potential 
and the effective $R$-dependent nuclear mass and the moment of inertia.
These three functions enter the nuclear Schr{\"o}dinger equation, 
which can be solved numerically for an arbitrary energy level. 
As a test of the presented perturbative theory, 
we perform  calculations of all 301 rovibrational levels of
H$_2$ molecule. We find an excellent $0.1\%$ agreement
with the accurate nonadiabatic corrections for states with the angular momentum
$J\leq 10$, which were obtained by Wolniewicz in \cite{Wol95},
and present for the first time results for states with $J>10$.

\section{The adiabatic approximation}
The total wave function $\phi$ is the solution of the stationary Schr\"odinger equation 
\begin{equation}
[H-E]\,|\phi\rangle = 0\,, \label{01}
\end{equation}
with the Hamiltonian 
\begin{equation}
H = H_{\rm el} + H_{\rm n}\,, \label{02}
\end{equation}
split into the electronic and nuclear parts.  In the electronic Hamiltonian
$H_{\rm el}$ 
\begin{equation}
H_{\rm el} = -\sum_{a}\frac{\nabla^2_a}{2\,m_{\rm e}} + V \label{03}
\end{equation}
nuclear masses are, by definition, set to infinity, and
the potential $V$ includes all the Coulomb interactions
with fixed positions $\vec R_A$ of nuclei. The nuclear Hamiltonian 
involves kinetic energies of all nuclei
\begin{equation}\label{ETn}
H_{\rm n} = -\sum_A\frac{\nabla^2_{\!R_A}}{2\,M_A}\,.
\end{equation}
The separation of center of mass motion and the choice
of the reference frame depends on the particular molecule.
For example, for a diatomic molecule
in the space fixed reference frame attached 
to the geometrical center of two nuclei $H_{\rm n}$ takes the form
\begin{equation}
H_{\rm n} = - \frac{\nabla^2_{\!R}}{2\,\mu_{\rm n}}
           - \frac{\nabla^2_{\rm el}}{2\,\mu_{\rm n}} 
           - \biggl(\frac{1}{M_B}-\frac{1}{M_A}\biggr)\,
             \vec\nabla_R\cdot\vec\nabla_{\rm el}\,,\label{EHn}
\end{equation}
where 
\begin{equation}
\vec\nabla_{\rm el}\equiv\frac{1}{2}\sum_a\vec\nabla_a\,,
\end{equation}
$\vec R = \vec R_{AB} = \vec R_A-\vec R_B$, and $1/\mu_{\rm n}=1/M_A+1/M_B$ 
is the nuclear reduced mass. The last term in Eq.~(\ref{EHn}) vanishes for 
homonuclear diatomic molecules.
 
In the adiabatic approximation the total wave function of an arbitrary  molecule 
\begin{equation}
\phi_{\rm a}(\vec r,\vec R) = \phi_{\rm el}(\vec r) \; \chi(\vec R) \label{04}
\end{equation}
is represented as a product of the electronic wave function $\phi_{\rm el}$
and the nuclear wave function $\chi$. We note, that $\phi_{\rm el}$
depends implicitly on the nuclear coordinates $\vec R$.
The electronic wave function obeys the clamped nuclei electronic 
Schr\"odinger equation 
\begin{equation}
\bigl[H_{\rm el}-{\cal E}_{\rm el}(\vec R)\bigr]\,|\phi_{\rm el}\rangle = 0, \label{05}
\end{equation} 
while the nuclear wave function is a solution to the Schr\"odinger equation 
in the effective potential generated by electrons
\begin{equation}
\bigl[ H_{\rm n} +{\cal E}_{\rm a}(\vec R)+{\cal E}_{\rm el}(\vec R)-E_{\rm a}\bigr]\,
|\chi\rangle = 0\,, \label{06}
\end{equation} 
where 
\begin{equation}
{\cal E}_{\rm a}(\vec R) = \bigl\langle\phiel|H_{\rm n}|\phiel\bigr\rangle_{\rm el}\,.
\label{ea}
\end{equation}
For the diatomic molecule the nuclear radial equation reads
\begin{eqnarray}
\biggl[-\frac{1}{2\,R^2}\,\frac{\partial}{\partial R}\,
\frac{R^2}{\mu_{\rm n}}\,\frac{\partial}{\partial R}\,
+\frac{J\,(J+1)}{2\,\mu_{\rm n}\,R^2}&& 
\nonumber \\ 
+ {\cal E}_{\rm a}(R)
+{\cal E}_{\rm el}(R)-E_{\rm a}\biggr]\,\chi_J(R) &=& 0\,, \label{09}
\end{eqnarray}
where $J$ is the rotational quantum number.

\section{Perturbative formalism}
The total wave function
\begin{equation}
\phi = \phi_{\rm a} + \delta\phi_{\rm na} = \phi_{\rm el}\,\chi + \delta\phi_{\rm na}
\label{07}
\end{equation}
is the sum of the adiabatic solution and a nonadiabatic correction. 
The nonadiabatic correction $\delta\phi_{\rm na}$ is  decomposed into two parts
\begin{equation}\label{10}
\delta\phi_{\rm na} = \phi_{\rm el}\,\delta\chi + \delta'\phi_{\rm na}\,,
\end{equation}
which  obey the following orthogonality conditions
\begin{eqnarray}
\langle\delta'\phi_{\rm na}|\phi_{\rm el}\rangle_{\rm el} &=& 0\,,\label{11}\\
\langle\delta\chi|\chi\rangle &=& 0\,.\label{12} 
\end{eqnarray}
The last equation means that normalization of $\phi$ is of the form
\begin{equation}
\langle\phi_{\rm el}\,\chi|\phi\rangle = 1\,.
\end{equation}
The total energy 
\begin{equation}
E = E_{\rm a}+\delta E_{\rm na}
\end{equation}
is the sum of the adiabatic energy $E_{\rm a}$ and the nonadiabatic 
correction $\delta E_{\rm na}$. Using above definitions we proceed
with the derivation of perturbative formulae.

The starting point is the Schr\"odinger equation (\ref{01}) with
the Hamiltonian $H$, the wave function $\phi$, and the energy $E$ 
decomposed into adiabatic and nonadiabatic parts
\begin{equation}
\bigl[(H_{\rm el}-{\cal E}_{\rm el})+
({\cal E}_{\rm el}+H_{\rm n}-E_{\rm a}-\delta E_{\rm na})]
|\phi_{\rm el}\,(\chi+\delta\chi)+\delta'\phi_{\rm na}\rangle = 0.
\label{star2}
\end{equation}
One rewrites this equation to the form
\begin{eqnarray}
({\cal E}_{\rm el}-H_{\rm el})|\delta'\phi_{\rm na}\rangle &=& 
({\cal E}_{\rm el}+H_{\rm n}-E_{\rm a}-\delta E_{\rm na})
\nonumber \\ &&
|\phi_{\rm el}\,(\chi+\delta\chi)+\delta'\phi_{\rm na}\rangle
\end{eqnarray}
and, since $\delta'\phi_{\rm na}$ is orthogonal to $\phi_{\rm el}$, 
Eq.~(\ref{11}), the formal solution 
\begin{eqnarray}
|\delta'\phi_{\rm na}\rangle &=& \frac{1}{({\cal E}_{\rm el}-H_{\rm el})'}
\bigl[ H_{\rm n}|\phi_{\rm el}\,(\chi+\delta\chi)\rangle
\nonumber \\ &&
+({\cal E}_{\rm el}+H_{\rm n}-E_{\rm a}-\delta E_{\rm na})
|\delta'\phi_{\rm na}\rangle\bigr]\label{dpsi},
\end{eqnarray}
is obtained,
where the prime in the denominator denotes subtraction of the reference
state from the Hamiltonian inversion. When $\delta\chi$ and 
$\delta'\phi_{\rm na}$ on the right hand side are neglected,
Eq. (\ref{dpsi}) becomes the leading nonadiabatic correction to the wave function.
In the next step one takes Eq.~(\ref{star2}) and multiplies it
from the left by $\langle\phi_{\rm el}|$
\begin{equation}
\langle\phi_{\rm el}|{\cal E}_{\rm el}+H_{\rm n}-E_{\rm a}-\delta E_{\rm na}
|\phi_{\rm el}\,(\chi+\delta\chi)+\delta'\phi_{\rm na}\rangle_{\rm el} = 0.
\end{equation}
Since $\chi$ satisfies Eq.~(\ref{06}) the above can be simplified to
\begin{eqnarray}
&&({\cal E}_{\rm el}+{\cal E}_{\rm a}
+H_{\rm n}-E_{\rm a})|\delta\chi\rangle
\nonumber \\&=&
\delta E_{\rm na}|\chi+\delta\chi\rangle
-\langle\phi_{\rm el}|H_{\rm n}|\delta'\phi_{\rm na}\rangle_{\rm el} 
\label{star3}
\end{eqnarray}
and due to Eq.~(\ref{12}) the solution is
\begin{eqnarray}
|\delta\chi\rangle &=& \frac{1}{\left(E_{\rm a}-{\cal E}_{\rm el}
-{\cal E}_{\rm a}-H_{\rm n}\right)'}
\nonumber \\ &&
\bigl(\langle\phi_{\rm el}|H_{\rm n}|\delta'\phi_{\rm na}\rangle_{\rm el}
-\delta E_{\rm na}|\chi+\delta\chi\rangle\bigr)\label{50}.
\end{eqnarray}
In the last step, one takes Eq.~(\ref{star3}),
multiplies it from the left by $\langle\chi|$, and obtains
\begin{equation}
\delta E_{\rm na} = 
\langle\phi_{\rm el}\,\chi|H_{\rm n}|\delta'\phi_{\rm na}\rangle\label{51}.
\end{equation}
The set of recursive equations (\ref{dpsi}),~(\ref{50}), and (\ref{51})
forms the perturbative expansion of the wave functions
$\delta'\phi_{\rm na}$, $\delta \chi$ and energy $\delta E_{\rm na}$.
For example, starting from (\ref{51}) one gets
\begin{eqnarray}
\delta E_{\rm na} &=& \langle\phi_{\rm el}\,\chi|H_{\rm n}
\frac{1}{({\cal E}_{\rm el}-H_{\rm el})'}
\Bigl[H_{\rm n}|\phi_{\rm el}\,(\chi+\delta\chi)\rangle
\nonumber \\ &&
+({\cal E}_{\rm el}+H_{\rm n}-E_{\rm a}-\delta E_{\rm na})
|\delta'\phi_{\rm na}\rangle\Bigr] \label{perturb},
\end{eqnarray}
which is the sum of the leading, Eq.~(\ref{Ena}),
and the higher order nonadiabatic correction, Eq.~(\ref{ho}).
This perturbative expansion in general assumes that 
${\cal E}_{\rm el}+H_{\rm n}-E_{\rm a}$
is small with respect to the electronic excitation energy.
It is not always true, especially for rovibrational levels close to
the dissociation threshold. In spite of this fact, we claim that
each power of ${\cal E}_{\rm el}+H_{\rm n}-E_{\rm a}$ 
in these particular matrix elements is at least of the order
$\mathcal{O}(\sqrt{m_{\rm e}/\mu_{\rm n}})$, which we demonstrate in next 
sections
for the leading terms $\delta^{(2)} E_{\rm na}$ and $\delta^{(3)} E_{\rm na}$
of the nonadiabatic perturbative expansion.
 
\subsection{Second-order nonadiabatic corrections}

In the leading order of perturbative treatment of nonadiabatic effects
one has
\begin{eqnarray}
|\delta'\phi_{\rm na}\rangle &=& 
\frac{1}{({\cal E}_{\rm el}-H_{\rm el})'}\,
H_{\rm n}\,|\phi_{\rm el}\,\chi\rangle,\label{16}\\
|\delta\chi\rangle &=& \frac{1}{\left(E_{\rm a}- 
{\cal E}_{\rm el}-{\cal E}_{\rm a}-H_{\rm n} \right)'}
\bigl\langle\phi_{\rm el}\bigl|H_{\rm n}\bigr|
\delta'\phi_{\rm na}\bigr\rangle_{\rm el}\,,
\nonumber \\ \label{21}\\
\delta^{(2)} E_{\rm na} &=&  
\biggl\langle\phi_{\rm el}\,\chi\biggl|
H_{\rm n}\,\frac{1}{({\cal E}_{\rm el}-H_{\rm el})'}\,
H_{\rm n}\,\biggr|\phi_{\rm el}\,\chi\biggr\rangle.
\label{Ena}
\end{eqnarray}
The general formula (\ref{Ena}), following  \cite{PK08b},
can be readily rearranged to a more practical form. 
From now on we consider the homonuclear diatomic two-electron molecule and
separate out electronic matrix elements from the nuclear ones 
\begin{eqnarray}
\delta^{(2)} E_{\rm na} &=& \int d^3R\left[
\chi^\star\,\chi\biggl\langle H_{\rm n}\,\phi_{\rm el}
\biggl|\frac{1}{({\cal E}_{\rm el}-H_{\rm el})'}\biggr| 
H_{\rm n}\,\phi_{\rm el}\biggr\rangle_{\!\rm el}\right.\nonumber \\&&
-\frac{\chi^\star\;\nabla_{\!R}^i\chi}{\mu_{\rm n}}\,
\biggl\langle H_{\rm n}\,\phi_{\rm el}
\biggl|\frac{1}{({\cal E}_{\rm el}-H_{\rm el})'}\biggr|
\nabla_{\!R}^i\phi_{\rm el}\biggr\rangle_{\!\rm el} \label{27}\\&&
-\frac{\nabla_{\!R}^i\chi^\star\;\chi}{\mu_{\rm n}}\,
\biggl\langle \nabla_{\!R}^i\,\phi_{\rm el}
\biggl|\frac{1}{({\cal E}_{\rm el}-H_{\rm el})'}\biggr|
H_{\rm n}\,\phi_{\rm el}\biggr\rangle_{\!\rm el}\nonumber \\&&
+\left.\frac{\nabla_{\!R}^i\chi^\star\,\nabla_{\!R}^j\chi}{\mu_{\rm n}^2}\,
\biggl\langle\nabla_{\!R}^i\phi_{\rm el}
\biggl|\frac{1}{({\cal E}_{\rm el}-H_{\rm el})'}\biggr|
\nabla_{\!R}^j\phi_{\rm el}\biggr\rangle_{\!\rm el}\right]\nonumber\\
&\equiv& \int d^3R\,\Bigl[ \chi^\star\,\chi\;{\mathcal U}(R) 
- \nabla_{\!R}^i\,\bigl[\chi^\star\,\chi\bigr]\,{\mathcal V}^i(R) \nonumber \\
&&\hspace*{11ex}
+\nabla_{\!R}^i\,\chi^\star\;\nabla_{\!R}^j\,\chi
\,{\mathcal W}^{ij}(R)\Bigr]\,,\label{Enaw}
\end{eqnarray}
where the last equation is the definition of potentials ${\mathcal U},
{\mathcal V}^i$ and ${\mathcal W}^{ij}$.
For the $\Sigma$ electronic state $\phi_{\rm el}$
\begin{eqnarray}
{\mathcal V}^i &=& n^i\,{\mathcal V}\,,\\
{\mathcal W}^{ij} &=& n^i\,n^j\,{\mathcal W}_\| 
+ (\delta^{ij}-n^i\,n^j)\,{\mathcal W}_\perp\,,
\end{eqnarray}
where $\vec n = \vec R/R$, hence
\begin{eqnarray}
\delta^{(2)} E_{\rm na} &=& \int d^3R\,\biggl\{ \chi^\star \chi\;\delta {\cal E}_{\rm na}(R)
\nonumber \\ &&
+n^i\,n^j\,\nabla_{\!R}^i\,\chi^\star\;\nabla_{\!R}^j\,\chi
\,{\mathcal W}_\|(R)
\nonumber \\ &&
+(\delta^{ij}-n^i\,n^j)\,\nabla_{\!R}^i\,\chi^\star\;\nabla_{\!R}^j\,\chi
\,{\mathcal W}_\perp(R)
\biggr\}\,.\quad\label{Enaw2}
\end{eqnarray}
The function
\begin{equation}
\delta {\cal E}_{\rm na}(R) =
\mathcal{U}(R) +\biggl(\frac{2}{R}+\frac{\partial}{\partial R}\biggr)\mathcal{V}(R) \label{43}
\end{equation}
is the nonadiabatic correction to the adiabatic energy curve 
${\cal E}_{\rm el}(R)+{\cal E}_{\rm a}(R)$ and pseudopotentials $\cal U, V, W$
are: 
\begin{eqnarray}
{\mathcal U}(R) &=&\biggl\langle H_{\rm n}\,\phi_{\rm el}
\biggl|\frac{1}{({\cal E}_{\rm el}-H_{\rm el})'}\biggr| 
H_{\rm n}\,\phi_{\rm el}\biggr\rangle_{\!\rm el}\,,\\
{\mathcal V}(R) &=&\frac{1}{\mu_{\rm n}}\,\biggl\langle H_{\rm n}\,\phi_{\rm el}
\biggl|\frac{1}{({\cal E}_{\rm el}-H_{\rm el})'}\biggr|
\vec n\cdot\vec\nabla_{\!R}\phi_{\rm el}\biggr\rangle_{\!\rm el}\,,\label{EVR}\\
{\mathcal W}_\|(R) &=&\frac{1}{\mu_{\rm n}^2}\,
\biggl\langle\vec n\cdot\vec\nabla_{\!R}\phi_{\rm el}
\biggl|\frac{1}{({\cal E}_{\rm el}-H_{\rm el})'}\biggr|
\vec n\cdot\vec\nabla_{\!R}\phi_{\rm el}\biggr\rangle_{\!\rm el}\,,
\nonumber\\ \\
{\mathcal W}_\perp(R) &=&\frac{1}{\mu_{\rm n}^2}\,
\frac{(\delta^{ij}-n^i\,n^j)}{2}\nonumber \\ &&\times
\biggl\langle\nabla^i_{\!R}\phi_{\rm el}
\biggl|\frac{1}{({\cal E}_{\rm el}-H_{\rm el})'}\biggr|
\nabla^j_{\!R}\phi_{\rm el}\biggr\rangle_{\!\rm el}\,.\label{EWT}
\end{eqnarray}
In order to simplify the nonadiabatic correction of Eq. (\ref{Enaw2}), 
one notes that the nuclear wave function $\chi$ has a definite angular momentum
\begin{equation}
\chi = \chi_{Jm}(\vec R) = \chi_J(R)\,Y_{Jm}(\vec n),
\end{equation}
where $Y_{Jm}$ are spherical harmonics, thus
the nonadiabatic correction can be  transformed to the form
\begin{eqnarray}
\delta^{(2)} E_{\rm na} &=& \int R^2\,dR\,\biggl\{ 
\chi_J'{}^{2}\,{\mathcal W}_\|(R)\label{Enaw3} \\ &&\hspace*{-10ex}
+\chi_J^2\,
\biggl[\delta{\cal E}_{\rm na}(R)
+\frac{J\,(J+1)}{R^2}\,{\mathcal W}_\perp(R)
\biggr]\biggr\}\,.\nonumber
\end{eqnarray}
Let us note, that
Eq.~(\ref{Enaw3}) can also be expressed in terms of an expectation value of
an effective nonadiabatic Hamiltonian $\delta H_{\rm na}$
\begin{equation}
\delta^{(2)} E_{\rm na} = \langle\chi_J|\delta H_{\rm na}|\chi_J\rangle\,, \label{30}
\end{equation}
where
\begin{eqnarray}
\delta H_{\rm na} &=& -\frac{1}{R^2}\,\frac{\partial}{\partial R}\,R^2\,\mathcal{W}_\|(R)\,
\frac{\partial}{\partial R}+\frac{J\,(J+1)}{R^2}\,{\mathcal W}_\perp(R)
\nonumber \\&& +\delta {\cal E}_{\rm na}(R).\label{EHna}
\end{eqnarray}
Now, the nonadiabatic correction $\delta\chi$ of Eq.~(\ref{21})
can be conveniently rewritten in terms of $\delta H_{\rm na}$
\begin{equation}
|\delta\chi_J\rangle = \frac{1}{\bigl(E_{\rm a}- {\cal E}_{\rm el}
-{\cal E}_{\rm a}-H_{\rm n} \bigr)'}\,\delta H_{\rm na}|\chi_J\rangle\,,
\end{equation}
where it is understood, that the derivatives with respect to electronic variables
of the function $\chi$ do vanish. 

\subsection{Third-order nonadiabatic correction}

The third order nonadiabatic correction of Eq.~(\ref{perturb}) is
\begin{eqnarray}
\delta^{(3)}E_{\rm na} &=&\biggl\langle\phi_{\rm el}\,\chi\biggl|
H_{\rm n}\,\frac{1}{({\cal E}_{\rm el}-H_{\rm el})'}\,
(H_{\rm n}+{\cal E}_{\rm el}-E_{\rm a})
\nonumber \\ &&
\times\frac{1}{({\cal E}_{\rm el}-H_{\rm el})'}\,
H_{\rm n}\,\biggr|\phi_{\rm el}\,\chi\biggr\rangle
\nonumber \\ \nonumber \\&&
+\biggl\langle\phi_{\rm el}\,\chi\biggl|
H_{\rm n}\,\frac{1}{({\cal E}_{\rm el}-H_{\rm el})'}\,
H_{\rm n}\,\biggr|\phi_{\rm el}\,\delta\chi\biggr\rangle,\label{ho}
\end{eqnarray}
where $\delta \chi$ is given in Eq.~(\ref{21}). 
Let us split this sum into two parts accordingly
\begin{equation}
\delta^{(3)}E_{\rm na} = \delta^{(3)}E'_{\rm na} + \delta^{(3)}E''_{\rm na}\,.
\end{equation}
While the first part $\delta^{(3)}E'_{\rm na}$
involves many terms which are negligible, since they include the third power
of $\mu_n$ in the denominator,
the dominating $\mathcal{O}(\mu_n^{-2})$ term is
\begin{eqnarray}
\delta^{(3)}E'_{\rm na} &=& \frac{1}{\mu_{\rm n}^2}\,
\left\langle\nabla^i_R\phi_{\rm el}\,\nabla^i_R\chi\left|
\frac{1}{({\cal E}_{\rm el}-H_{\rm el})'}
(H_{\rm n}+{\cal E}_{\rm el}-E_{\rm a})\right.\right.\nonumber \\&&
\left.\left.\times\frac{1}{({\cal E}_{\rm el}-H_{\rm el})'}
\right|\nabla^j_R\phi_{\rm el}\,\nabla^j_R\chi\right\rangle+\mathcal{O}(\mu_n^{-3})
\\&\approx&
-\int d^3R\;\vec n\cdot\vec\nabla_R(\chi^*\,\chi)\,\delta\mathcal{V}(R)
+\mathcal{O}(\mu_n^{-3})\,,
\label{Ed3E}
\end{eqnarray}
where
\begin{align}\label{EdV}
\delta\mathcal{V}(R) =&\frac{1}{2\,\mu_{\rm n}^2}\frac{\partial {\cal E}_{\rm el}}
{\partial R}\\ & \times \nonumber
\left\langle\vec n\cdot\vec\nabla_R\phi_{\rm el}\left|
\frac{1}{[({\cal E}_{\rm el}-H_{\rm el})']^2}
\right|\vec n\cdot\vec\nabla_R\phi_{\rm el}\right\rangle_\mathrm{el} \,,
\end{align}
and this correction is included into $\delta {\cal E}_{\rm na}$ of
Eq.~(\ref{43}), which becomes now
\begin{equation}
\delta\mathcal{E}_\mathrm{na}(R) =
\mathcal{U}(R) +\biggl(\frac{2}{R}+\frac{\partial}{\partial R}\biggr)
[\mathcal{V}(R)+\delta\mathcal{V}(R)]\,. \label{dena}
\end{equation}
The second term $\delta^{(3)}E''_{\rm na}$ can be obtained from the 
nonperturbative solution of the nuclear equation with the nonadiabatic 
Hamiltonian $\delta H_{\rm na}$ from Eq.~(\ref{EHna}). 
Namely, for the states with the rotational quantum number $J$, 
$\delta^{(3)}E''_{\rm na}$ takes the form
\begin{eqnarray}
\delta^{(3)}E''_{\rm na} &=& 
\langle\chi_J|\delta H_{\rm na}|\delta\chi_J\rangle \label{42} \\ &=& 
\left\langle\chi_J\left|\delta H_{\rm na}\,
\frac{1}{\bigl(E_{\rm a}- {\cal E}_{\rm el}- {\cal E}_{\rm a} -H_{\rm n}\bigr)'}\, 
\delta H_{\rm na}\right|\chi_J\right\rangle\,, \nonumber
\end{eqnarray}
which is interpreted as a second order correction due to $\delta H_{\rm na}$.

\subsection{Nuclear equation with effective potentials}
Instead of calculating the second order correction Eq.~(\ref{30}) and the third 
order correction of Eq.~(\ref{42}), it is more appropriate to solve 
nonperturbatively the corresponding equation
\begin{equation}
\bigl[H_{\rm n} + {\cal E}_{\rm el}(R) + {\cal E}_{\rm a}(R) 
+ \delta H_{\rm na}\bigr]\,\chi_J = E\,\chi_J \,,\label{nonper}
\end{equation}
where the total energy $E$ is
\begin{equation}
E = E_{\rm a} + \delta^{(2)}E_{\rm na} + \delta^{(3)}E_{\rm na} +\ldots 
\end{equation}
and this is done in this work.  Moreover, from comparison of subsequent terms 
of Eq.~(\ref{EHna}) with those of Eq.~(\ref{09}), one concludes 
that the first term can be interpreted as an $R$-dependent correction
to the nuclear reduced mass $\mu_{\rm n}$,
\begin{equation}
\frac{1}{2\,\mu_{\|}(R)} \equiv \frac{1}{2\,\mu_{\rm n}} + {\mathcal W}_\|(R),
\label{ERdmu}
\end{equation}
whereas the second term---as an $R$-dependent correction
to the inverse of the moment of inertia $\mu_\perp(R)\,R^2$
\begin{equation}
\frac{1}{2\,\mu_\perp(R)}\equiv \frac{1}{2\,\mu_{\rm n}} + {\mathcal W}_\perp(R).
\label{ERdmi}
\end{equation}
With these newly defined functions one can write the radial equation as
\begin{eqnarray}
&&\biggl[-\frac{1}{2\,R^2}\,\frac{\partial}{\partial R}\,
\frac{R^2}{\mu_{\|}(R)}\,\frac{\partial}{\partial R}\,
+\frac{J\,(J+1)}{2\,\mu_\perp(R) R^2}\, 
+\mathcal{Y}(R)\biggr]\,\chi_J(R) 
\nonumber \\&&= E\,\chi_J(R) \,,\label{nonperw}
\end{eqnarray}
where 
\begin{equation}\label{EY}
\mathcal{Y}(R) ={\cal E}_{\rm el}(R)+{\cal E}_{\rm a}(R)+\delta {\cal E}_{\rm na}(R) 
\end{equation}
forms the effective nonadiabatic potential.

\subsection{Asymptotics of the effective masses}
\label{subsec:Asymp}

The adiabatic correction ${\cal E}_{\rm a}(R)$, Eq.~(\ref{ea}), 
and the nonadiabatic
correction $\delta\mathcal{E}_\mathrm{na}(R)$, Eq.~(\ref{43}), 
do not vanish at large internuclear distances. For example,
for the large atomic separation in the hydrogen molecule, 
${\cal E}_{\rm a}(R)$ and $\delta\mathcal{E}_\mathrm{na}(R)$ are equal to
$m_{\rm e}/m_{\rm p}$ and $-(m_{\rm e}/m_{\rm p})^2$, respectively,
which corresponds to the first terms in the expansion of the atomic
reduced mass $\mu=(1/m_\mathrm{p}+1/m_\mathrm{e})^{-1}$
in the electron-nucleus mass ratio
\begin{equation}\label{Emas}
1-\frac{\mu}{m_{\rm e}} =\frac{m_{\rm e}/m_{\rm p}}{1+m_{\rm e}/m_{\rm p}}
= \frac{m_{\rm e}}{m_{\rm p}}
-\left(\frac{m_{\rm e}}{m_{\rm p}}\right)^2+\left(\frac{m_{\rm e}}{m_{\rm p}}\right)^3-\dots
\end{equation}
Large $R$ asymptotics of the pseudopotentials
${\mathcal W}_\|(R)$ and ${\mathcal W}_\perp(R)$ are
equal to $-m_{\rm e}/m_{\rm p}^2$, which is related to the change in 
Eqs.~(\ref{ERdmu}) and (\ref{ERdmi}) of the reduced nuclear mass $\mu_{\rm n}$ 
to the reduced mass $\mu_{\rm A}=(m_\mathrm{p}+m_\mathrm{e})/2$
of two hydrogen atoms :
\begin{eqnarray}
\frac{1}{2\,\mu_{\|}(\infty)} &=& \frac{1}{2\,\mu_{\perp}(\infty)}=
\frac{1}{2\,\mu_{\rm A}} \nonumber \\ &=&
 \frac{1}{m_{\rm p}+m_{\rm e}} = 
\frac{1}{m_{\rm p}}\biggl(1-\frac{m_{\rm e}}{m_{\rm p}}+\cdots\biggr).
\label{Amas}
\end{eqnarray}

\subsection{Evaluation of the wave function derivatives}

The electronic matrix elements in Eq.~(\ref{27}) involve 
multiple differentiation of the electronic 
wave function with respect to the internuclear distance $R$, 
which is difficult to calculate directly. 
Therefore, following Ref. \cite{PK08b}, 
we rewrite these terms to a more convenient form, 
where differentiation is taken of the Coulomb potential, namely
\begin{eqnarray}
\vec \nabla_{\!R}\phi_{\rm el} &=&\frac{1}{({\cal E}_{\rm el}-H_{\rm el})'}\,
\vec\nabla_{\!R}(V)\,\phi_{\rm el}\,,\label{ENRf}\\
\nabla^2_{\!R}\phi_{\rm el} &=&\frac{1}{({\cal E}_{\rm el}-H_{\rm el})'}\,
\biggl\{\nabla_{\!R}^2(V)\,\phi_{\rm el}
 +2\,\vec\nabla_{\!R}(V-{\cal E}_{\rm el})\label{ELRf}\\ && 
\frac{1}{({\cal E}_{\rm el}-H_{\rm el})'}\,
\vec\nabla_{\!R} (V)\,\phi_{\rm el}\biggr\}
+\phi_{\rm el}\,\langle \phi_{\rm el}|\nabla^2_{\!R}|\phi_{\rm el}\rangle_{\rm el},
\nonumber
\end{eqnarray}
The derivatives of the potential $V$ 
\begin{equation}
V =-\frac{1}{r_{1A}}-\frac{1}{r_{1B}}-\frac{1}{r_{2A}}-\frac{1}{r_{2B}}
+\frac{1}{r_{12}}+ \frac{1}{R_{AB}} \,,
\end{equation}
are the following
\begin{eqnarray}
\vec\nabla_{\!R}(V) &=& \frac{1}{2}\,\biggl(
-\frac{\vec r_{1A}}{r_{1A}^3}+\frac{\vec r_{1B}}{r_{1B}^3}
-\frac{\vec r_{2A}}{r_{2A}^3}+\frac{\vec r_{2B}}{r_{2B}^3}
\biggr)-\frac{\vec R}{R^3}\,,\nonumber \\ \\
\nabla_{\!R}^2(V) &=& \pi\,\bigl[
\delta(\vec r_{1A}) + \delta(\vec r_{1B}) +
\delta(\vec r_{2A}) + \delta(\vec r_{2B})\bigr] 
\nonumber \\&&
-4\,\pi\,\delta(\vec R)\,,\label{EnrsH}
\end{eqnarray}
and the matrix elements with these operators are readily evaluated.
The presence of the Dirac delta operators in Eq.~(\ref{ELRf}) may potentially
decrease the accuracy of the evaluation of those quantities which
contain $\nabla^2_{\!R}\phi_{\rm el}$. If we note that 
\begin{equation}\label{EnesH}
\nabla^2_{\rm el}(V)=\pi\left[\delta(\vec{r}_{1A})+\delta(\vec{r}_{1B})
+\delta(\vec{r}_{2A})+\delta(\vec{r}_{2B})\right],
\end{equation}
we can get rid of Dirac deltas by a simple rearrangement
of the nuclear part of the Hamiltonian $H_\mathrm{n}$ to the form
\begin{equation}
H_\mathrm{n}=-\frac{1}{2\mu_n}\,\left(\nrs-\nabla^2_{\rm el}\right)
-\frac{1}{\mu_n}\,\nabla^2_{\rm el}\,.
\end{equation}
The difference in parenthesis collects the terms
of Eq.~(\ref{EnrsH}) and (\ref{EnesH}) which cancel out, up to the negligible
$-4\,\pi\,\delta(\vec R)$ term, so that we can write it down as
\begin{eqnarray}
&&\left(\nrs-\nabla^2_{\rm el}\right)\phiel=
\label{Enrmns} \\ &&
\frac{2}{({\cal E}_{\rm el}-H_{\rm el})'}
\nr{}(V-{\cal E}_{\rm el})
\Hinvp\nr{}(V)\,\phiel\nonumber\\&&
-\frac{2}{({\cal E}_{\rm el}-H_{\rm el})'}\,\vec\nabla_{\rm el}(V)
\Hinv\vec\nabla_{\rm el}(V)\,\phiel+\lambda\phiel
\nonumber
\end{eqnarray}
with some constant $\lambda$. In this way the Dirac delta 
terms are eliminated at the expense of employing additional basis sets for 
evaluation of the last resolvent in Eq.~(\ref{Enrmns}).

\section{Numerical calculations}

In order to form the radial nonadiabatic equation (\ref{nonperw}),
apart from the clamped nuclei energy ${\cal E}_{\rm el}(R)$ and
the adiabatic correction $\mathcal{E}_\mathrm{a}(R)$,
it is necessary to know the pseudopotentials ${\cal U}(R), 
{\cal V}(R) +\delta{\cal V}(R), {\cal W}_\|(R)$, and ${\cal W}_\perp(R)$.
The evaluation of these functions is the main 
numerical task described here. Calculations were performed at 80 points 
including long and very short internuclear distances $R$. 
At each point several electronic wave functions of different symmetry were
generated. All these functions were represented as linear
expansions of properly symmetrized two-electron basis functions. 
The basis functions were taken in the form of exponentially correlated 
Gaussians (ECG)
\begin{eqnarray}\label{EECG}
\psi_k(\br_1,\br_2)&=&(1+\hat{P}_{12})(1\pm\hat\imath)\,\Xi_k\\&\times&
\exp{\left[-\sum_{i,j=1}^2 A_{ij,k}(\br_i-\bs_{i,k})(\br_j-\bs_{j,k})\right]},
\nonumber
\end{eqnarray}
where the matrices $\mathbf{A}_k$ and vectors $\bs_k$ contain nonlinear
parameters, 5 per basis function, to be variationally optimized. 
The antisymmetry projector
$(1+\hat{P}_{12})$ ensures singlet symmetry, the spatial projector 
$(1\pm\hat\imath)$---the gerade ($+$) or ungerade ($-$) symmetry, 
and the $\Xi_k$ prefactor enforces
$\Sigma$ states when equal to 1, or $\Pi$ states when equal to $y_i$---the 
perpendicular Cartesian component of the electron coordinate.

Six different basis sets have been prepared to represent the variety of the
electronic wave functions. To ensure high accuracy of the potentials,
the basis sets have been variationally optimized with respect to pertinent 
goal functions according to the specification in Table~\ref{Tb}.

\begin{table}[!hbt]
\caption{\label{Tb} Goal functions used in optimization of the basis sets.}
\renewcommand{\arraystretch}{2.0}
\begin{tabular*}{0.48\textwidth}{c@{\extracolsep{\fill}}lcl}
\hline
Label & Size & Symmetry & Goal function \\
\hline
A & 600 & $\Sigma_g$ & ${\cal E}_{\rm el}$ \\
B & $600^\dag$ & $\Sigma_g$ & $\ds\Bigl\langle\nabla_R^\|(V)
\frac{1}{({\cal E}_{\rm el}-H_{\rm el})'}\nabla_R^\|(V)\Bigr\rangle$\\
C & 1200 & $\Pi_g$ & $\ds\Bigl\langle\nabla_R^\perp(V)
\frac{1}{{\cal E}_{\rm el}-H_{\rm el}}\nabla_R^\perp(V)\Bigr\rangle$ \\
D & $600^\dag$ & $\Sigma_g$ & $\ds\Bigl\langle\nabla_{\rm el}^2
\frac{1}{({\cal E}_{\rm el}-H_{\rm el})'}\nabla_{\rm el}^2\Bigr\rangle$ \\
E & 600 & $\Sigma_u$ & $\ds\Bigl\langle\nabla_{\rm el}^\|(V)
\frac{1}{{\cal E}_{\rm el}-H_{\rm el}}\nabla_{\rm el}^\|(V)\Bigr\rangle$ \\
F & 600 & $\Pi_u$ & $\ds\Bigl\langle\nabla_{\rm el}^\perp(V)
\frac{1}{{\cal E}_{\rm el}-H_{\rm el}}\nabla_{\rm el}^\perp(V)\Bigr\rangle$ \\[1ex]
\hline
\end{tabular*}

$^\dag$ Optimized along with the fixed basis A.
\end{table}

Particular goal functions have been chosen to reflect the contents of the 
expression the basis set is to be used for.
The first basis (labeled A), composed of 600 ECG functions (\ref{EECG}), was 
employed to expand the X$^1\Sigma_g^+$ electronic ground state wave function 
$\phi_{\mathrm{el}}$. Their nonlinear parameters were optimized variationally 
with respect to the clamped nuclei energy with the target accuracy of the order 
of a fraction of microhartree. The bases B, C, and D were intended for
evaluation of the resolvents present in Eqs.~(\ref{ENRf}) and (\ref{Enrmns}).
The two $\Sigma_g^+$ bases (B and D) were optimized in the presence
of the basis A: the first 600 terms
were taken from $\phi_{\mathrm{el}}$ wave function and their
nonlinear parameters were kept fixed during the optimization, only the remaining
600 terms were actually optimized. This ensures that
the internal wave function $\phi_{\rm el}$ is well represented at every step
of optimization. Then, the subtraction of the reference state, denoted by the $'$ 
symbol within the resolvent, was achieved by
orthogonalization of $\vec\nabla_{\!R}(V)\,|\phi_{\rm el}\rangle$
to the internal $|\phi_{\rm el}\rangle$. 
In the final calculations the three bases A, B, and D were assembled together
to form a 1800-term $\Sigma_g^+$ basis applied not only to evaluate
the pertinent resolvents but also 
to expand the external ground state function $\phi_{\mathrm{el}}$.
The two ungerade bases (E and F) were employed to evaluate
the resolvent and to form the components of the scalar product in the second term
of the right hand side of Eq.~(\ref{Enrmns}).

The adiabatic potential of the nuclear Schr{\"o}dinger equation (\ref{07}) was 
composed of the clamped nuclei energy, $\mathcal{E}_{\mathrm{el}}(R)$,
and the adiabatic correction $\mathcal{E}_{\mathrm{a}}(R)$.
For $\mathcal{E}_{\mathrm{el}}(R)$ we used the analytic potential constructed
by \L ach \cite{GL-thesis} on the basis of the energy points computed by Cencek from 
1200-term ECG wave functions \cite{Cen-pc} and Sims and Hagstrom from Hylleraas
wave functions \cite{SH06}. Their energy points where converged up to 13 significant digits.
The adiabatic correction ${\cal E}_{\rm a}(R)$ was
evaluated as an expectation value of the Hamiltonian $H_{\rm n}$, Eq.~(\ref{ea}),
\begin{eqnarray}\label{Ead}
{\cal E}_{\rm a}(R)&=&
-\frac{1}{2\mu_{\mathrm{n}}}\langle\phi_{\mathrm{el}}|\nabla_{\!R}^2
+ \nabla_{\rm el}^2|\phi_{\mathrm{el}}\rangle_{\mathrm{el}}\,.
\end{eqnarray}
To avoid the cumbersome differentiation of the electronic wave function 
with respect to the internuclear distance we replaced the expectation value
in the first term on the right
hand side of Eq.~(\ref{Ead}) by an equivalent expression 
\begin{equation}
\langle\phi_{\rm el}|\nabla_{\!R}^2|\phi_{\rm el}\rangle_{\rm el}=
-\langle\vec\nabla_{\!R}\phi_{\rm
el}|\vec\nabla_{\!R}\phi_{\rm el}\rangle_{\rm el}\,,
\end{equation}
which, with the help of Eq.~(\ref{ENRf}), can be further transformed to
\begin{equation}
\langle\phi_{\mathrm{el}}|\nabla_{\!R}^2|\phi_{\mathrm{el}}\rangle_{\mathrm{el}}
=-\left\langle\vec\nabla_{\!R}(V)
\frac{1}{\bigl[({\cal E}_{\rm el}-H_{\rm el})'\bigr]^2}
\vec\nabla_{\!R}(V)\right\rangle_{\rm el}.\label{ELRwe}
\end{equation}
The formula (\ref{ELRwe}), when evaluated with the optimized bases A, B, and C,
yields the adiabatic correction with an accuracy of at least 1 ppm. 
The adiabatic potential curve was then obtained by means of 10-point 
piecewise polynomial interpolation.

\begin{figure}[!ht]
\includegraphics[width=6.0cm,angle=-90]{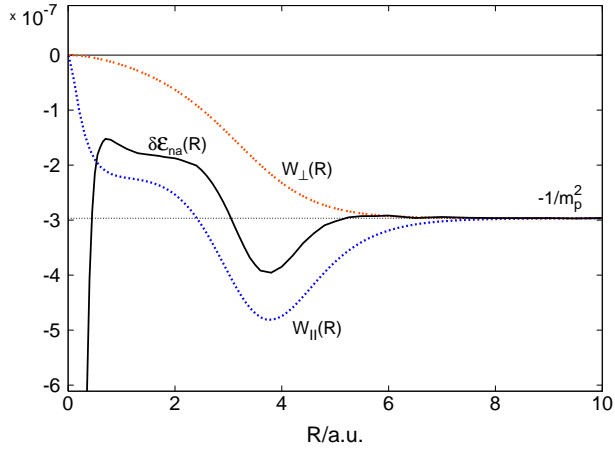}
\caption{\label{FRYWW} (Color online) The nonadiabatic potentials (in a.u.): 
$\delta\mathcal{E}_\mathrm{na}(R)$ (bolded, black), $\mathcal{W}_\parallel(R)$ 
(lower dotted, blue), and $\mathcal{W}_\perp(R)$ (upper dotted, red).
The potentials asymptote goes at $-1/m_\mathrm{p}^2=-2.966077\times 10^{-7}$.}
\end{figure}

The electronic matrix elements $\mathcal{U}$, $\mathcal{V}+\delta\mathcal{V}$, 
$\mathcal{W}_\|$, 
$\mathcal{W}_\perp$ entering Eq.~(\ref{Enaw2}) were evaluated with the 
ECG basis sets described above, yielding smooth functions of $R$.
Because for the highest vibrational levels the nuclear wave functions are
spread out and the contributions from larger internuclear distances are
non-negligible, the functions $\mathcal{U}(R)$, $\mathcal{V}(R)$, and
$\mathcal{W}(R)$ were represented by their asymptotic forms:
\begin{eqnarray}
\mathcal{U}(R) &\approx& u_0+u_6/R^{6}+u_8/R^{8},\nonumber \\
\mathcal{V}(R) &\approx& v_9/R^9+v_{11}/R^{11},\nonumber \\
\delta\mathcal{V}(R) &\approx& v_7/R^7+v_{0}/R^9,\nonumber \\ 
\mathcal{W_\|}(R) &\approx& w_{\| 0}+w_{\| 12}/R^{12}+w_{\| 14}/R^{14},\nonumber \\
\mathcal{W_\perp}(R) &\approx& w_{\perp 0}+w_{\perp 12}/R^{12}+w_{\perp 14}/R^{14},
\end{eqnarray}
subject to $u_0=w_{\| 0}=w_{\perp 0}=-(m_\mathrm{e}/m_\mathrm{p})^2$ restriction 
(in atomic units). The remaining, free parameters
$u_i$, $v_i$, and $w_i$ were determined by fitting the above functions to
the calculated points in the range of $\langle 6.0,10.0\rangle$ bohrs.
Because at distances $R>6$, the numerical precision of the potentials $\cal U$ 
and $\mathcal{V}$ was not high enough, we used lower $R$-values for the 
extrapolation.
At the origin $R=0$ all the potentials are finite with ${\cal V}\sim R$,
${\cal W_\|}\sim R^2$, and ${\cal W_\perp}\sim R^2$. Numerical results
for $\delta{\cal E}_{\rm na}$, ${\cal W_\|}$, and ${\cal W_\perp}$
are shown graphically in Fig. \ref{FRYWW}.

The nonadiabatic correction to energy levels can be computed perturbatively 
from Eq.~(\ref{Enaw3}) as has been done in our previous paper
\cite{PK08b}, or, in a more accurate way, the radial equation (\ref{nonper}) 
can be solved for the total nonadiabatic energy. This second method
is described shortly below. The first term including the second order derivative 
is rewritten to the more convenient form
\begin{eqnarray}
&&-\frac{1}{R^2}\,\frac{\partial}{\partial R}\,
R^2\,\biggl(\frac{1}{2\,\mu_{\rm n}}+{\mathcal W}_\|(R)\biggr)\,\frac{\partial}{\partial R}
\nonumber \\ 
&&= -\frac{1}{R}\,\frac{\partial}{\partial R}\,
\biggl(\frac{1}{2\,\mu_{\rm n}}+{\mathcal W}_\|(R)\biggr)\,\frac{\partial}{\partial R}\,R
+\frac{{\mathcal W}'_\|(R)}{R}\,.\quad
\end{eqnarray}
The resulting radial equation 
\begin{align}
&\frac{\partial}{\partial R}\,\frac{1}{\mu_\|(R)}\,  \nonumber
\frac{\partial}{\partial R}\,\eta_J(R)\\ \label{Ere}
&= -2\biggl[E-\mathcal{Y}(R)-\frac{{\mathcal W}'_\|(R)}{R}
-\frac{J\,(J+1)}{2\,\mu_\perp(R)\,R^2}\biggr]\,\eta_J(R),
\end{align}
is solved numerically for the function $\eta_J(R)=R\,\chi_J(R)$. 
We used the code developed by W. Johnson, described recently
in his book \cite{john}, and modified it to account for the dependence 
of the mass on the internuclear distance. 
In the calculations we used the following constants \cite{NIST}:
the proton mass $m_\mathrm{p}=1836.15267247\ m_\mathrm{e}$ and
the energy units conversion factor 1 hartree~$=219474.6313705$~cm$^{-1}$.

\section{Results and discussion}

\begin{figure}[!ht]
\includegraphics[width=6.0cm,angle=-90]{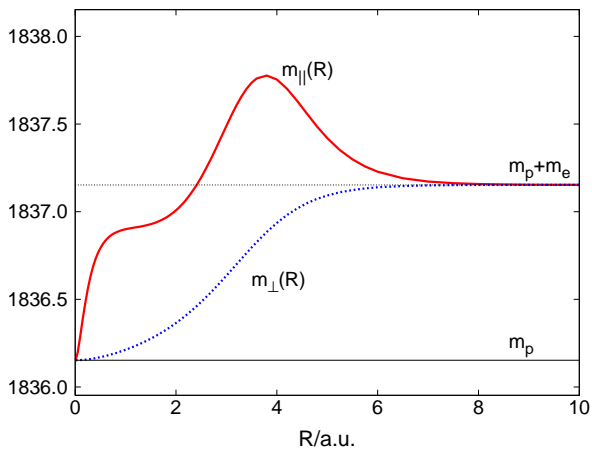}
\caption{\label{Fmp} (Color online) The $R$-dependence of the effective
nuclear masses $m_{\|}(R)$ and $m_{\perp}(R)$ (in a.u.).}
\end{figure}

In a molecule, the moving nuclei are 'coated' with electrons and the amount 
of additional mass carried by the nuclei changes with $R$. 
For a homonuclear molecule in a purely vibrational state, the effective
mass of the nucleus at given $R$, $m_{\|}(R)$, is just twice the 
reduced mass $\mu_{\|}(R)$ defined in Eq.~(\ref{ERdmu}). Analogously, 
for a rigid rotating molecule, the effective nuclear mass $m_{\perp}(R)$
is related to $\mu_{\perp}(R)$ of Eq.~(\ref{ERdmi}). 
Thus, their $R$-dependence can be determined explicitly from the potentials 
$\mathcal{W}_{\|}(R)$ and $\mathcal{W}_{\perp}(R)$, respectively.
Fig.~\ref{Fmp} illustrates the changes in the two effective nuclear masses
with the internuclear distance in H$_2$. The functions $m_{\|}(R)$ and 
$m_{\perp}(R)$ join smoothly the proton mass at the united atom limit
with the hydrogen atomic mass ($m_\mathrm{p}+m_\mathrm{e}$) at the separated
atoms limit. Interestingly, for $R\ge 2.41$ a.u.,
the effective mass $m_{\|}(R)$ is greater than the sum of proton
and electron masses, reaching $m_\mathrm{p}+1.6\,m_\mathrm{e}$ at the maximum 
located near $R=3.8$ a.u.

The radial equation (\ref{Ere}) has been solved for all bound states
with three versions of the potential $\mathcal{Y}(R)$: 
\begin{equation}
\begin{array}{ll}
\mathcal{Y}(R)={\cal E}_{\rm el}(R), &{\rm BO}\\
\mathcal{Y}(R)={\cal E}_{\rm el}(R)+{\cal E}_{\rm a}(R), & {\rm adiabatic} \\
\mathcal{Y}(R)={\cal E}_{\rm el}(R)+{\cal E}_{\rm a}(R)+\delta {\cal E}_{\rm na}(R),
& {\rm nonadiabatic}
\end{array}\nonumber
\end{equation}
yielding three sets of dissociation energies. 
The corresponding dissociation thresholds were $-1$ hartree 
in the BO approximation,
$-1+m_\mathrm{e}/m_\mathrm{p}$ hartree in adiabatic approximation, 
and $-1+m_\mathrm{e}/m_\mathrm{p}-\left(m_\mathrm{e}/m_\mathrm{p}\right)^2$ 
hartree in the nonadiabatic level 
of theory. The results are listed in Table~\ref{Tres}, where for each pair
of quantum numbers $v$ and $J$ three entries are given (in cm$^{-1}$): 
the BO dissociation energy, the adiabatic correction, 
and the nonadiabatic correction to the dissociation energy. 
Thus, the total nonrelativistic dissociation energy can be obtained by summing up 
all three entries.
The only exception is the state with $v=14$ and $J=4$, for which a nonadiabatic 
level lying just beneath the dissociation threshold has been predicted although
neither BO nor adiabatic bound states exist. The entry given for this state
is the energy separation from the nonadiabatic dissociation threshold.

Our nonadiabatic corrections agree very well 
with those computed by Wolniewicz \cite{Wol95} for rovibrational states of
$J\leq 10$. For all rotational states of the lowest vibrational level 
the difference is merely 0.0002 cm$^{-1}$ or less. In all the cases
the difference is smaller than $0.1\%$, which corresponds to the magnitude 
of the neglected higher order terms of relative order $\mathcal{O}(1/\mu_{\rm n})$. 
This is the first confirmation of the Wolniewicz's results \cite{Wol95} 
for $J>0$ rovibrational states and, simultaneously, a numerical validation 
of the nonadiabatic perturbation theory presented in this work. 
Let us note, that the nonadiabatic corrections to the
dissociation energy from Table~\ref{Tres} differ in sign from the corrections 
to the adiabatic energy of \cite{Wol95} and their absolute
values differ by the constant $1/m_\mathrm{p}^2=0.06509787$ cm$^{-1}$ 
corresponding to the asymptotic value of the nonadiabatic potential
$\delta\mathcal{E}_\mathrm{na}(R)$,  discussed in Subsec.~\ref{subsec:Asymp}.

\section{Summary}

We have presented the nonadiabatic perturbation theory applicable to 
any molecule in an arbitrary rovibrational state.
The leading nonadiabatic corrections for the diatomic molecule
are expressed in terms
of three $R$-dependent functions depicted in Fig.~\ref{FRYWW}: the nuclear 
reduced mass in Eq.~(\ref{ERdmu}), 
the moment of inertia in Eq.~(\ref{ERdmi}), and the correction Eq.~(\ref{dena})
to the adiabatic potential, which enter the radial Schr\"odinger
equation for nuclei Eq.~(\ref{nonper}). This equation can be solved
for an arbitrary molecular states by standard numerical methods \cite{john}. 
Although representation of the nonadiabatic correction by pseudopotentials
has been studied previously (see \cite{BM77,Kut07,JK08} and 
references therein), we have rigorously derived 
new expressions for ${\cal W}_\|$, ${\cal W}_\perp$, and
$\delta{\cal E}_{\rm na}$ functions of the internuclear distance,
which give nonadiabatic corrections with $O(m_{\rm e}/\mu_{\rm n})$ accuracy.
Moreover, we have performed explicit numerical calculations for the simplest 
example of H$_2$ molecule in order to verify the obtained perturbative
formulae. All the electronic matrix elements with differentiation
of the wave function over the internuclear distance were rewritten 
in a convenient form involving differentiation of the 
Coulomb potential. Such an approach enables achieving much higher numerical 
precision even for the well known adiabatic correction.
The final accuracy of all three nonadiabatic functions
is limited only by the neglected higher order terms, namely 
$\mathcal{O}(m_{\rm e}/\mu_{\rm n})$ relative to the leading order, and 
results for rovibrational levels agree within this 
uncertainty with former results of Wolniewicz \cite{Wol95}.
Having accurate nonrelativistic energies one can 
include relativistic and QED corrections, which become
significant for the states close to the dissociation threshold.
Relativistic corrections for the electronic ground state of H$_2$
are known accurately for a wide range of internuclear distances \cite{Wol93}.
Their large $R$ asymptotics, including QED corrections, is presently
investigated by the Jeziorski group \cite{BJ}. Combining all the knowledge
together would enable achieving at least an order of magnitude increase in
the precision of theoretical prediction for all molecular states of H$_2$.

Analogous calculations can be performed for D$_2$  and HD molecules.
It is worth noting, that in the former case there is no need to recompute
the pseudopotentials and only a proper rescaling due to different reduced
masses is required.  
The calculations on the latter system, due to difference in mass of the proton 
and deuteron, would involve additional correction to $\delta{\cal E}_{\rm na}$ 
coming from the last term of Eq.~(\ref{EHn}).
While this perturbative approach can be further extended to larger 
diatomic molecules, it would be more challenging to investigate
three-atomic molecules such as H$_3^+$---a system of great
astrophysical interest. We expect no principal difficulties in such calculations
except for much more increased demands in computer resources needed to perform
optimization of pertinent wave functions. 

 \section*{Acknowledgments}
We are indebted to L.~Wolniewicz for valuable comments.
Part of the computations has been performed in Pozna\'n\ Supercomputing
and Networking Center.

\clearpage
 \begin{table*}[!htb]
 \caption{\label{Tres}
The dissociation energy of the rovibrational states of H$_2$ (in cm$^{-1}$). 
For each pair of vibrational ($v$) and rotational ($J$) quantum numbers, three
entries are given: the BO dissociation energy, the adiabatic correction, 
and the nonadiabatic correction. The sum of the three numbers gives the total 
nonrelativistic dissociation energy of the ($v$, $J$) state. }
 \begin{ruledtabular}
 \begin{tabular}{c@{\extracolsep{\fill}}*{ 8}{w{5.4}}}
$v\backslash J$& \cent{ 0}  & \cent{ 1}  & \cent{ 2}  & \cent{ 3}  & \cent{ 4}  & \cent{ 5}  & \cent{ 6}  & \cent{ 7}    \\
 \hline
  0 & 36112.5927 & 35994.0372 & 35758.0143 & 35406.6660 & 34943.1216 & 34371.4040 & 33696.3152 & 32923.3094   \\
  0 &     5.7711 &     5.8348 &     5.9611 &     6.1481 &     6.3931 &     6.6924 &     7.0418 &     7.4367   \\
  0 &     0.4339 &     0.4406 &     0.4539 &     0.4739 &     0.5005 &     0.5338 &     0.5736 &     0.6200   \\
 \\
  1 & 31949.1892 & 31836.5534 & 31612.3337 & 31278.6001 & 30838.3755 & 30295.5437 & 29654.7385 & 28921.2187   \\
  1 &     7.1740 &     7.2311 &     7.3443 &     7.5119 &     7.7310 &     7.9983 &     8.3096 &     8.6605   \\
  1 &     1.2704 &     1.2761 &     1.2874 &     1.3044 &     1.3271 &     1.3555 &     1.3896 &     1.4294   \\
 \\
  2 & 28021.4345 & 27914.5889 & 27701.9154 & 27385.4159 & 26968.0130 & 26453.4605 & 25846.2343 & 25151.4125   \\
  2 &     8.3336 &     8.3840 &     8.4839 &     8.6314 &     8.8240 &     9.0583 &     9.3304 &     9.6358   \\
  2 &     2.0271 &     2.0318 &     2.0414 &     2.0557 &     2.0749 &     2.0989 &     2.1279 &     2.1618   \\
 \\
  3 & 24324.4498 & 24223.3066 & 24022.0055 & 23722.4851 & 23327.5744 & 22840.9061 & 22266.8106 & 21610.1998   \\
  3 &     9.2420 &     9.2855 &     9.3714 &     9.4982 &     9.6632 &     9.8633 &    10.0945 &    10.3525   \\
  3 &     2.7088 &     2.7127 &     2.7206 &     2.7326 &     2.7485 &     2.7686 &     2.7929 &     2.8214   \\
 \\
  4 & 20855.2072 & 20759.7245 & 20569.7150 & 20287.0583 & 19914.4976 & 19455.5548 & 18914.4286 & 18295.8815   \\
  4 &     9.8890 &     9.9251 &     9.9966 &    10.1017 &    10.2378 &    10.4019 &    10.5902 &    10.7985   \\
  4 &     3.3187 &     3.3219 &     3.3283 &     3.3379 &     3.3508 &     3.3670 &     3.3868 &     3.4100   \\
 \\
  5 & 17612.7145 & 17522.9036 & 17344.2107 & 17078.4626 & 16728.3253 & 16297.2225 & 15789.2368 & 15209.0018   \\
  5 &    10.2629 &    10.2915 &    10.3476 &    10.4298 &    10.5356 &    10.6618 &    10.8049 &    10.9606   \\
  5 &     3.8576 &     3.8599 &     3.8647 &     3.8719 &     3.8816 &     3.8938 &     3.9087 &     3.9262   \\
 \\
  6 & 14598.2891 & 14514.2237 & 14346.9982 & 14098.3935 & 13771.0098 & 13368.1874 & 12893.9129 & 12352.7160   \\
  6 &    10.3525 &    10.3729 &    10.4130 &    10.4712 &    10.5449 &    10.6313 &    10.7268 &    10.8271   \\
  6 &     4.3218 &     4.3232 &     4.3262 &     4.3306 &     4.3365 &     4.3440 &     4.3531 &     4.3639   \\
 \\
  7 & 11815.9479 & 11737.7779 & 11582.3242 & 11351.3310 & 11047.3466 & 10673.6480 & 10234.1517 &  9733.3170   \\
  7 &    10.1486 &    10.1606 &    10.1839 &    10.2169 &    10.2573 &    10.3020 &    10.3476 &    10.3901   \\
  7 &     4.7029 &     4.7031 &     4.7037 &     4.7044 &     4.7055 &     4.7067 &     4.7081 &     4.7095   \\
 \\
  8 &  9272.9561 &  9200.9278 &  9057.7445 &  8845.1263 &  8565.5890 &  8222.3732 &  7819.3619 &  7360.9906   \\
  8 &     9.6483 &     9.6517 &     9.6577 &     9.6648 &     9.6707 &     9.6726 &     9.6670 &     9.6500   \\
  8 &     4.9838 &     4.9824 &     4.9796 &     4.9752 &     4.9692 &     4.9614 &     4.9516 &     4.9393   \\
 \\
  9 &  6980.5984 &  6915.0837 &  6784.9238 &  6591.8323 &  6338.3203 &  6027.6323 &  5663.6708 &  5250.9187   \\
  9 &     8.8590 &     8.8537 &     8.8422 &     8.8232 &     8.7945 &     8.7533 &     8.6963 &     8.6196   \\
  9 &     5.1364 &     5.1325 &     5.1246 &     5.1124 &     5.0958 &     5.0743 &     5.0472 &     5.0138   \\
 \\
 10 &  4955.2699 &  4896.8118 &  4780.7739 &  4608.8912 &  4383.7137 &  4108.5514 &  3787.4126 &  3424.9438   \\
 10 &     7.8021 &     7.7882 &     7.7597 &     7.7151 &     7.6525 &     7.5691 &     7.4616 &     7.3260   \\
 10 &     5.1160 &     5.1082 &     5.0925 &     5.0684 &     5.0354 &     4.9926 &     4.9386 &     4.8720   \\
 \\
 11 &  3220.0418 &  3169.4253 &  3069.1017 &  2920.8723 &  2727.3991 &  2492.1652 &  2219.4373 &  1914.2414   \\
 11 &     6.5140 &     6.4918 &     6.4469 &     6.3777 &     6.2821 &     6.1573 &     5.9996 &     5.8045   \\
 11 &     4.8566 &     4.8429 &     4.8151 &     4.7725 &     4.7140 &     4.6379 &     4.5418 &     4.4224   \\
 \\
 12 &  1806.9489 &  1765.3260 &  1683.0605 &  1562.1049 &  1405.3733 &  1216.7368 &  1001.0464 &   764.2101   \\
 12 &     5.0372 &     5.0068 &     4.9451 &     4.8505 &     4.7203 &     4.5506 &     4.3360 &     4.0685   \\
 12 &     4.2657 &     4.2429 &     4.1965 &     4.1254 &     4.0271 &     3.8984 &     3.7345 &     3.5281   \\
 \\
 13 &   760.3903 &   729.5279 &   668.9437 &   580.9383 &   469.0254 &   338.0600 &   194.5461 &    47.4825   \\
 13 &     3.3933 &     3.3526 &     3.2697 &     3.1417 &     2.9631 &     2.7253 &     2.4123 &     1.9886   \\
 13 &     3.2221 &     3.1850 &     3.1095 &     2.9927 &     2.8294 &     2.6113 &     2.3232 &     1.9317   \\
 \\
 14 &   141.7951 &   124.7523 &    92.3077 &    48.0033   \\
 14 &     1.5343 &     1.4739 &     1.3479 &     1.1416   \\
 14 &     1.5847 &     1.5226 &     1.3933 &     1.1825 & 0.0887^\dag \\
 \\
 \end{tabular}
 \end{ruledtabular}

$^\dag$ This state appears as a resonance in BO and adiabatic approximations.
The entry is a dissociation energy of this nonadiabatic level.\hfill\
 \end{table*}

\setcounter{table}{1}
 \begin{table*}[!htb]
 \caption{continued}
 \begin{ruledtabular}
 \begin{tabular}{c@{\extracolsep{\fill}}*{ 8}{w{5.4}}}
$v\backslash J$& \cent{ 8}  & \cent{ 9}  & \cent{10}  & \cent{11}  & \cent{12}  & \cent{13}  & \cent{14}  & \cent{15}    \\
 \hline
  0 & 32058.3583 & 31107.8190 & 30078.3065 & 28976.5781 & 27809.4302 & 26583.6105 & 25305.7462 & 23982.2860   \\
  0 &     7.8718 &     8.3418 &     8.8411 &     9.3642 &     9.9055 &    10.4596 &    11.0212 &    11.5852   \\
  0 &     0.6728 &     0.7322 &     0.7981 &     0.8704 &     0.9491 &     1.0343 &     1.1260 &     1.2240   \\
 \\
  1 & 28100.7394 & 27199.4234 & 26223.6397 & 25179.8929 & 24074.7260 & 22914.6378 & 21706.0160 & 20455.0847   \\
  1 &     9.0458 &     9.4605 &     9.8991 &    10.3562 &    10.8264 &    11.3045 &    11.7850 &    12.2629   \\
  1 &     1.4749 &     1.5263 &     1.5834 &     1.6464 &     1.7154 &     1.7903 &     1.8713 &     1.9584   \\
 \\
  2 & 24374.5496 & 23521.5539 & 22598.5709 & 21611.8782 & 20567.7938 & 19472.5994 & 18332.4783 & 17153.4682   \\
  2 &     9.9697 &    10.3271 &    10.7026 &    11.0910 &    11.4869 &    11.8851 &    12.2802 &    12.6671   \\
  2 &     2.2008 &     2.2449 &     2.2942 &     2.3488 &     2.4089 &     2.4744 &     2.5456 &     2.6225   \\
 \\
  3 & 20876.4463 & 20071.2651 & 19200.6029 & 18270.5391 & 17287.1990 & 16256.6838 & 15185.0135 & 14078.0862   \\
  3 &    10.6327 &    10.9300 &    11.2393 &    11.5554 &    11.8729 &    12.1865 &    12.4908 &    12.7804   \\
  3 &     2.8543 &     2.8917 &     2.9337 &     2.9804 &     3.0320 &     3.0885 &     3.1500 &     3.2167   \\
 \\
  4 & 17605.1247 & 16847.7045 & 16029.3982 & 15156.1198 & 14233.8415 & 13268.5282 & 12266.0890 & 11232.3425   \\
  4 &    11.0221 &    11.2561 &    11.4954 &    11.7348 &    11.9690 &    12.1923 &    12.3995 &    12.5849   \\
  4 &     3.4370 &     3.4677 &     3.5023 &     3.5409 &     3.5836 &     3.6304 &     3.6814 &     3.7365   \\
 \\
  5 & 14561.5917 & 13852.4141 & 13087.1119 & 12271.4764 & 11411.3755 & 10512.6972 &  9581.3093 &  8623.0355   \\
  5 &    11.1244 &    11.2915 &    11.4566 &    11.6146 &    11.7601 &    11.8873 &    11.9906 &    12.0642   \\
  5 &     3.9466 &     3.9699 &     3.9962 &     4.0254 &     4.0576 &     4.0926 &     4.1304 &     4.1704   \\
 \\
  6 & 11749.5653 & 11089.7678 & 10378.8781 &  9622.6208 &  8826.8275 &  7997.3920 &  7140.2429 &  6261.3368   \\
  6 &    10.9279 &    11.0242 &    11.1110 &    11.1829 &    11.2344 &    11.2597 &    11.2528 &    11.2071   \\
  6 &     4.3763 &     4.3903 &     4.4058 &     4.4226 &     4.4406 &     4.4592 &     4.4779 &     4.4957   \\
 \\
  7 &  9176.0486 &  8567.6048 &  7913.5170 &  7219.5226 &  6491.5173 &  5735.5270 &  4957.7045 &  4164.3543   \\
  7 &    10.4250 &    10.4474 &    10.4524 &    10.4343 &    10.3876 &    10.3060 &    10.1831 &    10.0114   \\
  7 &     4.7108 &     4.7118 &     4.7119 &     4.7108 &     4.7076 &     4.7012 &     4.6902 &     4.6724   \\
 \\
  8 &  6852.1602 &  6298.1569 &  5704.5851 &  5077.3199 &  4422.4828 &  3746.4465 &  3055.8806 &  2357.8507   \\
  8 &     9.6172 &     9.5638 &     9.4847 &     9.3741 &     9.2260 &     9.0337 &     8.7892 &     8.4833   \\
  8 &     4.9241 &     4.9053 &     4.8820 &     4.8528 &     4.8160 &     4.7692 &     4.7092 &     4.6313   \\
 \\
  9 &  4794.3657 &  4299.4465 &  3771.9991 &  3218.2507 &  2644.8437 &  2058.9184 &  1468.2878 &   881.7751   \\
  9 &     8.5188 &     8.3891 &     8.2250 &     8.0203 &     7.7678 &     7.4587 &     7.0814 &     6.6193   \\
  9 &     4.9730 &     4.9232 &     4.8624 &     4.7880 &     4.6962 &     4.5819 &     4.4375 &     4.2515   \\
 \\
 10 &  3026.3806 &  2597.5217 &  2144.7391 &  1675.0460 &  1196.2612 &   717.3517 &   249.1650   \\
 10 &     7.1577 &     6.9514 &     6.7003 &     6.3963 &     6.0280 &     5.5786 &     5.0193   \\
 10 &     4.7905 &     4.6909 &     4.5694 &     4.4200 &     4.2343 &     3.9995 &     3.6930   \\
 \\
 11 &  1582.3694 &  1230.4417 &   866.0734 &   498.2452 &   138.1669   \\
 11 &     5.5658 &     5.2756 &     4.9227 &     4.4892 &     3.9416   \\
 11 &     4.2752 &     4.0939 &     3.8689 &     3.5852 &     3.2148   \\
 \\
 12 &   513.3908 &   257.4738 &     8.3253   \\
 12 &     3.7358 &     3.3166 &     2.7633   \\
 12 &     3.2682 &     2.9354 &     2.4878   \\
 \\
 \end{tabular}
 \end{ruledtabular}
 \end{table*}
\setcounter{table}{1}
 \begin{table*}[!htb]
 \caption{continued}
 \begin{ruledtabular}
 \begin{tabular}{c@{\extracolsep{\fill}}*{ 8}{w{5.4}}}
$v\backslash J$& \cent{16}  & \cent{17}  & \cent{18}  & \cent{19}  & \cent{20}  & \cent{21}  & \cent{22}  & \cent{23}    \\
 \hline
  0 & 22619.4577 & 21223.2371 & 19799.3296 & 18353.1603 & 16889.8728 & 15414.3355 & 13931.1527 & 12444.6818   \\
  0 &    12.1465 &    12.7005 &    13.2425 &    13.7681 &    14.2729 &    14.7525 &    15.2029 &    15.6196   \\
  0 &     1.3286 &     1.4396 &     1.5572 &     1.6814 &     1.8122 &     1.9497 &     2.0939 &     2.2451   \\
 \\
  1 & 19167.8656 & 17850.1517 & 16507.4922 & 15145.1868 & 13768.2885 & 12381.6135 & 10989.7585 &  9597.1235   \\
  1 &    12.7332 &    13.1911 &    13.6317 &    14.0502 &    14.4420 &    14.8023 &    15.1262 &    15.4086   \\
  1 &     2.0516 &     2.1512 &     2.2570 &     2.3693 &     2.4881 &     2.6134 &     2.7453 &     2.8838   \\
 \\
  2 & 15941.4274 & 14702.0131 & 13440.6714 & 12162.6372 & 10872.9422 &  9576.4337 &  8277.7999 &  6981.6067   \\
  2 &    13.0406 &    13.3957 &    13.7272 &    14.0300 &    14.2988 &    14.5281 &    14.7122 &    14.8446   \\
  2 &     2.7052 &     2.7938 &     2.8884 &     2.9890 &     3.0955 &     3.2080 &     3.3263 &     3.4502   \\
 \\
  3 & 12941.6497 & 11781.2863 & 10602.4104 &  9410.2757 &  8209.9952 &  7006.5729 &  5804.9493 &  4610.0657   \\
  3 &    13.0501 &    13.2944 &    13.5077 &    13.6843 &    13.8181 &    13.9026 &    13.9307 &    13.8939   \\
  3 &     3.2886 &     3.3657 &     3.4479 &     3.5352 &     3.6272 &     3.7237 &     3.8240 &     3.9274   \\
 \\
  4 & 10172.9970 &  9093.6453 &  7999.7724 &  6896.7794 &  5790.0229 &  4684.8764 &  3586.8202 &  2501.5762   \\
  4 &    12.7427 &    12.8672 &    12.9520 &    12.9905 &    12.9756 &    12.8991 &    12.7513 &    12.5207   \\
  4 &     3.7957 &     3.8587 &     3.9253 &     3.9948 &     4.0665 &     4.1392 &     4.2113 &     4.2803   \\
 \\
  5 &  7643.6478 &  6648.8775 &  5644.4446 &  4636.1112 &  3629.7676 &  2631.5640 &  1648.1214 &   686.8842   \\
  5 &    12.1016 &    12.0965 &    12.0417 &    11.9291 &    11.7496 &    11.4918 &    11.1414 &    10.6779   \\
  5 &     4.2124 &     4.2556 &     4.2990 &     4.3412 &     4.3801 &     4.4126 &     4.4344 &     4.4383   \\
 \\
  6 &  5366.6735 &  4462.3376 &  3554.5745 &  2649.9163 &  1755.3887 &   878.8631 &    29.7112   \\
  6 &    11.1158 &    10.9712 &    10.7645 &    10.4851 &    10.1192 &     9.6476 &     9.0402   \\
  6 &     4.5115 &     4.5235 &     4.5291 &     4.5249 &     4.5055 &     4.4629 &     4.3836   \\
 \\
  7 &  3361.9951 &  2557.4749 &  1758.1694 &   972.3301 &   209.7400   \\
  7 &     9.7827 &     9.4869 &     9.1113 &     8.6376 &     8.0373   \\
  7 &     4.6449 &     4.6038 &     4.5430 &     4.4532 &     4.3193   \\
 \\
  8 &  1660.0067 &   970.9237 &   300.7647   \\
  8 &     8.1037 &     7.6332 &     7.0443   \\
  8 &     4.5286 &     4.3911 &     4.2015   \\
 \\
  9 &   309.8952   \\
  9 &     6.0447   \\
  9 &     4.0047   \\
 \\
 \end{tabular}
 \end{ruledtabular}
 \end{table*}
\setcounter{table}{1}
 \begin{table*}[!htb]
 \caption{continued}
 \begin{ruledtabular}
 \begin{tabular}{c@{\extracolsep{\fill}}*{ 8}{w{5.4}}}
$v\backslash J$& \cent{24}  & \cent{25}  & \cent{26}  & \cent{27}  & \cent{28}  & \cent{29}  & \cent{30}  & \cent{31}    \\
 \hline
  0 & 10959.0542 &  9478.2014 &  8005.8846 &  6545.7303 &  5101.2727 &  3676.0048 &  2273.4444 &   897.2218   \\
  0 &    15.9981 &    16.3340 &    16.6221 &    16.8572 &    17.0329 &    17.1423 &    17.1769 &    17.1261   \\
  0 &     2.4031 &     2.5682 &     2.7404 &     2.9199 &     3.1066 &     3.3006 &     3.5019 &     3.7105   \\
 \\
  1 &  8207.9415 &  6826.3161 &  5456.2676 &  4101.7929 &  2766.9437 &  1455.9363 &   173.3131   \\
  1 &    15.6440 &    15.8264 &    15.9492 &    16.0045 &    15.9831 &    15.8734 &    15.6600   \\
  1 &     3.0290 &     3.1807 &     3.3389 &     3.5032 &     3.6734 &     3.8488 &     4.0283   \\
 \\
  2 &  5692.3452 &  4414.4956 &  3152.6140 &  1911.4560 &   696.1630   \\
  2 &    14.9184 &    14.9254 &    14.8559 &    14.6980 &    14.4356   \\
  2 &     3.5795 &     3.7135 &     3.8516 &     3.9923 &     4.1340   \\
 \\
  3 &  3426.9550 &  2260.8735 &  1117.5022   \\
  3 &    13.7827 &    13.5850 &    13.2850   \\
  3 &     4.0327 &     4.1379 &     4.2403   \\
 \\
  4 &  1435.3165 &   395.0112   \\
  4 &    12.1920 &    11.7435   \\
  4 &     4.3427 &     4.3926   \\
 \\
 \end{tabular}
 \end{ruledtabular}
 \end{table*}

\end{document}